**Bohmian version of a 2-state quantum system**

*Matthew Dickau,* matt.dickau@gmail.com

I construct a Bohmian version of a 2-state quantum system, where in addition to the quantum state, the system is characterized by a definite physical state which changes stochastically under the guidance of the quantum state. I argue that the probabilities produced by this process are well-defined, contrary to claims in [2] about a related model.

## Introduction

In an article [1] "Realism and the quantum-mechanical two-state oscillator," Hardy, Home, et. al. construct a model for a simple quantum system in which the system always has a determinate configuration, which changes stochastically in a way that reproduces the predictions of quantum mechanics. In their model, the probability rates for the system to jump from one configuration to the other rely on a further property of the system, a "clock variable", which is affected by measurements of the configuration. They note that this is analogous to the way that definite particle positions in Bohmian mechanics are guided by the wavefunction.

In reply [2] "Why quantum mechanics cannot be formulated as a Markov process," Gillespie criticizes Hardy et. al.'s model for (among other problems) having probability rates that diverge to infinity at certain times, arguing that this leads to probabilities that are not well-defined. Hardy et. al. respond in [3] and Gillespie has a final reply in [4]. In a recent discussion between Jacob Barandes and David Albert [5], Barandes references Gillespie's criticism and applies it to attempted extensions of Bohmian mechanics to quantum field theory [6][7] which have a stochastic element; this is also relevant to other models similar to [1], such as the one in [8] featuring a Bohmian particle hopping between 2 waveguides.

Here I contend that, contrary to [2], a Bohmian version of any 2-state quantum system does have well-defined probabilities, removing this argument against extensions of Bohmian mechanics with stochastic evolution of the physical state.

## The Bohmian Model

Consider a system which can be in two states, which I will label $S = 0$ and $S = 1$, with dynamics causing an oscillation between the states. Use basis states for the qubit:

$$|0\rangle = \begin{bmatrix} 1 \\ 0 \end{bmatrix}, \qquad |1\rangle = \begin{bmatrix} 0 \\ 1 \end{bmatrix}$$

And the Hamiltonian (shifted from the Hamiltonian in Refs. [1] and [2] by an irrelevant constant energy, and using units where $\hbar = 1$):

$$\hat{H} = \frac{1}{2}\omega\begin{bmatrix} 0 & 1 \\ 1 & 0 \end{bmatrix} = \frac{1}{2}\omega(|0\rangle\langle 1| + |1\rangle\langle 0|)$$

The solution to the Schrodinger equation for a given initial quantum state is:

$$i\frac{d|\psi\rangle}{dt} = \hat{H}|\psi\rangle\,, \qquad |\psi(t)\rangle = e^{-i\hat{H}t}|\psi(0)\rangle = \begin{bmatrix} \cos(\omega t/2) & -i\sin(\omega t/2) \\ -i\sin(\omega t/2) & \cos(\omega t/2) \end{bmatrix}|\psi(0)\rangle$$

For a Bohmian version of this system, in addition to the quantum state $|\psi(t)\rangle$, the oscillator also has a determinate physical state $S(t) \in \{0,1\}$, and it stochastically jumps back and forth between the two possible states with rates that are controlled by the quantum state.

Since the evolution of the quantum state is deterministic, we only need to know it at the initial time to know it at all times, barring interruptions in its evolution due to measurement effects. In the Bohmian model, measurement of the value of $S$ will cause collapse of the quantum state of the system, which can be treated as resetting the initial condition and the time variable (to $t = 0$ and either $S(0) = 0, |\psi(0)\rangle = |0\rangle$, or $S(0) = 1, |\psi(0)\rangle = |1\rangle$, depending on the value of $S$ at the time of measurement). More precisely, it causes effective collapse of the conditional quantum state of the subsystem, conditioned on the physical state of the measurement apparatus, while the combined quantum state of the measurement apparatus and subsystem evolves unitarily.

Note that measurements of the state of the system made relative to a different basis, e.g., the plus/minus basis:

$$|\pm\rangle = \frac{1}{\sqrt{2}}(|0\rangle \pm |1\rangle)$$

are contextual in the Bohmian model, depending on details of the measurement apparatus, since there is no physical configuration corresponding to this value. Such a measurement would collapse the quantum state to either $|+\rangle$ or $|-\rangle$ while stochastically guiding the physical state $S$ to equal 0 or 1 each with 50% probability (or whatever the appropriate probabilities are, for measurement in a different basis).

We can analyze what the stochastic jump rates should be to match the quantum probabilities. Use the Schrodinger equation to find how the probabilities change over time:

$$y_a(t) = |\langle a|\psi(t)\rangle|^2 = \langle\psi(t)|\hat{P}_a|\psi(t)\rangle\,, \qquad \hat{P}_a = |a\rangle\langle a|$$

$$\frac{dy_a(t)}{dt} = i\langle\psi(t)|\hat{H}\hat{P}_a|\psi(t)\rangle - i\langle\psi(t)|\hat{P}_a\hat{H}|\psi(t)\rangle$$

Inserting the identity operator $\hat{I} = \hat{P}_0 + \hat{P}_1$ before $\hat{H}$ in the first term on the right side and after it in the second term, we can rearrange these equations to read:

$$\frac{dy_0(t)}{dt} = -J(t)\,, \qquad \frac{dy_1(t)}{dt} = J(t)$$

Where the probability flux is defined by:

$$J(t) = \langle\psi(t)|\hat{J}|\psi(t)\rangle\,, \qquad \hat{J} = i\big(\hat{P}_0\hat{H}\hat{P}_1 - \hat{P}_1\hat{H}\hat{P}_0\big)$$

Define the "positive part" function $[x]^+ = x$ if $x > 0$, otherwise $[x]^+ = 0$. Then for any real number $x$, $x = [x]^+ - [-x]^+$, and we can rewrite:

$$\frac{dy_0(t)}{dt} = -[J(t)]^+ + [-J(t)]^+\,, \qquad \frac{dy_1(t)}{dt} = [J(t)]^+ - [-J(t)]^+$$

Finally, this can be rewritten by defining "jump rates" $W_{0\to1}$ and $W_{1\to0}$:

$$\frac{dy_0(t)}{dt} = -W_{0\to1}(t)y_0(t) + W_{1\to0}(t)y_1(t)\,, \qquad \frac{dy_1(t)}{dt} = W_{0\to1}(t)y_0(t) - W_{1\to0}(t)y_1(t)$$

$$W_{0\to1}(t) = \frac{[J(t)]^+}{y_0(t)}\,, \qquad W_{1\to0}(t) = \frac{[-J(t)]^+}{y_1(t)}$$

These equations can be heuristically interpreted as showing "probability fluid" flowing between the two states, with $W_{a\to b}(t)dt$ being the probability, in an infinitesimal time $dt$, that the system jumps to state $S = b$, given that it is in state $S = a$. Note that these rates can be modified by adding an equal positive flux $J'$ in both directions (balancing additional jumps from 0 to 1 with additional jumps from 1 to 0):

$$W_{0\to1}(t) = \frac{[J(t)]^+ + J'(t)}{y_0(t)}\,, \qquad W_{1\to0}(t) = \frac{[-J(t)]^+ + J'(t)}{y_1(t)}$$

However, this is not a rigorous derivation of the behaviour of the model; I will explore that in the next section.

The above analysis works for any Hamiltonian for the two-state system. For the form of the Hamiltonian given above, assuming the initial quantum state $|\psi(0)\rangle = |0\rangle$, the quantum state and probabilities are:

$$|\psi(t)\rangle = e^{-i\hat{H}t}|0\rangle = \cos(\omega t/2)\,|0\rangle - i\sin(\omega t/2)\,|1\rangle$$

$$y_0(t) = \cos^2(\omega t/2) = \frac{1}{2}(1 + \cos(\omega t))$$

$$y_1(t) = \sin^2(\omega t/2) = \frac{1}{2}(1 - \cos(\omega t))$$

While the probability flux and minimal jump rates (with $J' = 0$) are:

$$J(t) = \omega \sin(\omega t/2) \cos(\omega t/2) = \frac{\omega}{2} \sin(\omega t)$$

$$W_{0\to 1}(t) = \omega \frac{[\sin(\omega t)]^+}{1 + \cos(\omega t)}, \qquad W_{1\to 0}(t) = \omega \frac{[-\sin(\omega t)]^+}{1 - \cos(\omega t)}$$

Choosing $J' = \frac{\omega}{2}(\lambda - [-\sin(\omega t)]^+)$, with $\lambda > [-\sin(\omega t)]^+$, gives the general class of jump rates considered in [1]. Alternatively, choosing the limiting value $\lambda = [-\sin(\omega t)]^+$ in [1] gives these minimal jump rates.

**(Attempt at) A More Rigorous Derivation**

Given a known quantum state $|\psi(t)\rangle$ evolved by the Schrodinger equation from an initial state $|\psi(0)\rangle$, the stochastic process for the physical state $S(t)$ of the system is defined by conditional probabilities from earlier to later times, along with initial probabilities. In what follows, all probabilities are implicitly conditional on the known quantum state.

Here we are going to begin by assuming that probabilities for the stochastic process exist and have well-defined values, which we are trying to discover. If, in attempting to construct probabilities for this stochastic process, we inevitably run into a contradiction (such as probabilities below 0 or above 1), then Gillespie's criticisms will have been vindicated. But if we succeed in constructing probabilities without contradiction, then this kind of Bohmian process still appears to be possible.

Thus, assume there are single-time and multi-time probabilities for the stochastic process, such as $P(S(t) = a)$, $P(S(t_1) = a_1 \cap \ldots \cap S(t_n) = a_n)$, etc. In the multi-time case, we only assume the probabilities are defined for a finite number of conjuncts. Furthermore, we consider there to only ever be a finite number of instances of this system. Then since $S$ only has two possible values (and not infinitely many), it is reasonable to assume that events with probability zero never in fact occur for this process.

Conditional probabilities of the form $P(S(t') = b \mid S(t_1) = a_1 \cap \ldots \cap S(t_n) = a_n)$ for times $t_1, \ldots, t_n < t'$ are, in a sense, physically instantiated when $S$ has taken on the specified values $a_i$ at the specified times $t_i$, as the objective tendencies for the future behaviour of the process. However, if the conjunction $C = \bigcap_i \{S(t_i) = a_i\}$ has zero probability, probabilities conditioned on it are never physically instantiated – the stochastic process itself guarantees this – so nothing requires $P(S(t') = b \mid C)$ to be well-defined if $P(C) = 0$. So, we will assume that conditional probabilities $P(S(t') = b \mid C)$ exist if $P(C) > 0$, but not if $P(C) = 0$. Note that the conditional probabilities cannot be determined merely from the single-time probabilities; they require more information.

Probability axioms require that:

$$P(S(t') = b \mid C)P(C) = P(S(t') = b \cap C)$$

If $P(C) = 0$ then we can reasonably stipulate that $P(S(t') = b \cap C) = 0$ (and so also the product on the left-hand side equals zero), whereas if $P(C) > 0$, we can write:

$$P(S(t') = b \mid C) = \frac{P(S(t') = b \cap C)}{P(C)}$$

Thus, where they are required to be defined, the conditional probabilities obey the familiar ratio formula in terms of the "unconditioned" probabilities (more accurately, the probabilities that are only conditioned on the initial quantum state).

Now, the initial probabilities are special cases of the single-time probabilities, which are defined for $t \geq 0$:

$$p_a(t) \equiv P(S(t) = a)$$

We want to construct the conditional probabilities in such a way that these single-time probabilities always match the quantum probabilities, $p_a(t) = y_a(t)$, given the quantum equilibrium hypothesis for the initial probabilities:

$$p_a(0) = y_a(0)$$

Denote the (forward-directed) conditional probabilities for $0 \leq t \leq t'$:

$$p_{a\to b}(t,t') \equiv P(S(t') = b \mid S(t) = a) = \frac{P(S(t') = b \cap S(t) = a)}{P(S(t) = a)}$$

Again, note that, as a probability conditioned on $S(t) = a$, the value of $p_{a\to b}(t,t')$ need only be defined when $p_a(t) > 0$. However, we can still stipulate, consistent with the probability axioms, that the product $p_a(t)p_{a\to b}(t,t') = 0$ when $p_a(t) = 0$, even though $p_{a\to b}(t,t')$ may be undefined there (since $P(A)P(B|A) = P(A \cap B)$, and $P(A \cap B) = 0$ if $P(A) = 0$).

Note that we could also define the backwards-directed probabilities, for $0 \leq t \leq t'$:

$$q_{a\to b}(t,t') \equiv P(S(t) = a \mid S(t') = b) = \frac{P(S(t') = b \cap S(t) = a)}{P(S(t') = b)}$$

And find them from the forward-directed probabilities by Bayes' rule:

$$q_{a\to b}(t,t') = \frac{p_{a\to b}(t,t')p_a(t)}{p_b(t')}$$

However, we will not use these here.

We can connect the single-time probabilities to the conditional probabilities and the initial probabilities, via the law of total probability:

$$P(S(t)=a)=\sum_{c\in\{0,1\}} P(S(t)=a \mid S(0)=c)P(S(0)=c)$$

$$p_a(t)=\sum_{c\in\{0,1\}} p_{c\to a}(0,t)p_c(0)$$

So now we need to analyze the conditional probabilities to determine how to arrive at the desired relationship, $p_a(t)=y_a(t)$.

First, since $S=0$ and $S=1$ are mutually exclusive and exhaustive at every time, we of course have $p_1(t)=1-p_0(t)$, and, where they are defined, $p_{a\to 1}(t,t')=1-p_{a\to 0}(t,t')$.

It is also the case that:

$$p_{0\to 1}(t,t)=p_{1\to 0}(t,t)=0$$

Since these are probabilities for the state to be 0 and 1 simultaneously.

Now, another application of the law of total probability gives us, for $0\le t\le t'\le t''$:

$$P(S(t'')=b \mid S(t)=a)=\sum_{c\in\{0,1\}} P(S(t')=c \mid S(t)=a)P(S(t'')=b \mid S(t')=c\cap S(t)=a)$$

We will stipulate the Markov condition for this process: conditioned on the known quantum state, any probabilities conditioned on multiple past states reduce to the probability conditioned only on the most recent past state (where here $0\le t_1<\cdots<t_n\le t$):

$$P(S(t)=a \mid S(t_1)=a_1\cap\ldots\cap S(t_n)=a_n)=P(S(t)=a \mid S(t_n)=a_n)$$

Therefore, the previous statement from the law of total probability can be rewritten:

$$p_{a\to b}(t,t'')=\sum_{c\in\{0,1\}} p_{a\to c}(t,t')p_{c\to b}(t',t'')$$

Consider $0\le t_0\le t$ and $\delta t>0$ where both $p_0(t')>0$ and $p_1(t')>0$ for $t'$ in the interval $(t_0,t+\delta t)$. That is, the following derivation only considers intervals of time during which the single-time probabilities are strictly between 0 and 1. Then, by our assumptions, the conditional probabilities are well-defined, and:

$$p_{a\to 0}(t_0,t+\delta t)=p_{a\to 0}(t_0,t)p_{0\to 0}(t,t+\delta t)+p_{a\to 1}(t_0,t)p_{1\to 0}(t,t+\delta t)$$

$$p_{a\to 1}(t_0,t+\delta t)=p_{a\to 0}(t_0,t)p_{0\to 1}(t,t+\delta t)+p_{a\to 1}(t_0,t)p_{1\to 1}(t,t+\delta t)$$

These can be rewritten:

$$p_{a\to 0}(t_0, t+\delta t) = p_{a\to 0}(t_0, t)\big(1 - p_{0\to 1}(t, t+\delta t)\big) + p_{a\to 1}(t_0, t)p_{1\to 0}(t, t+\delta t)$$

$$p_{a\to 1}(t_0, t+\delta t) = p_{a\to 0}(t_0, t)p_{0\to 1}(t, t+\delta t) + p_{a\to 1}(t_0, t)\big(1 - p_{1\to 0}(t, t+\delta t)\big)$$

And rearranged:

$$p_{a\to 0}(t_0, t+\delta t) - p_{a\to 0}(t_0, t) = -p_{a\to 0}(t_0, t)p_{0\to 1}(t, t+\delta t) + p_{a\to 1}(t_0, t)p_{1\to 0}(t, t+\delta t)$$

$$p_{a\to 1}(t_0, t+\delta t) - p_{a\to 1}(t_0, t) = p_{a\to 0}(t_0, t)p_{0\to 1}(t, t+\delta t) - p_{a\to 1}(t_0, t)p_{1\to 0}(t, t+\delta t)$$

Divide by $\delta t$, and take the limit $\delta t \to 0^+$. Making the further assumption that $p_{a\to b}(t_0, t)$ is differentiable in its second argument, we get the <u>probability rate equations</u>:

$$\frac{\partial p_{a\to 0}(t_0, t)}{\partial t} = -W_{0\to 1}(t)p_{a\to 0}(t_0, t) + W_{1\to 0}(t)p_{a\to 1}(t_0, t)$$

$$\frac{\partial p_{a\to 1}(t_0, t)}{\partial t} = W_{0\to 1}(t)p_{a\to 0}(t_0, t) - W_{1\to 0}(t)p_{a\to 1}(t_0, t)$$

Where (for $a \neq b$, and using the fact that $p_{a\to b}(t, t) = 0$ in that case):

$$W_{a\to b}(t) \equiv \lim_{\delta t\to 0^+} \frac{p_{a\to b}(t, t+\delta t)}{\delta t} = \lim_{\delta t\to 0^+} \frac{p_{a\to b}(t, t+\delta t) - p_{a\to b}(t, t)}{\delta t} = \left.\frac{\partial p_{a\to b}(t, t')}{\partial t'}\right|_{t'=t}$$

We can similarly consider $\delta t < 0$ and perform the same derivation, switching the temporal order of $t$ and $t + \delta t$, and take the limit $\delta t \to 0^-$. We arrive at equations of the same form, except that now we have:

$$W_{a\to b}(t) \equiv -\lim_{\delta t\to 0^-} \frac{p_{a\to b}(t+\delta t, t)}{\delta t} = -\left.\frac{\partial p_{a\to b}(t', t)}{\partial t'}\right|_{t'=t}$$

From either definition, we have an argument that $W_{a\to b}(t)$ may become undefined when $p_a(t) = 0$ (though, strictly speaking, this is a trivial statement, in that we have only defined $W_{a\to b}(t)$ where $p_a(t) > 0$). From the first definition, $W_{a\to b}(t)$ may be undefined if $p_a(t) = 0$ because the underlying conditional probability, $p_{a\to b}(t, t')$, may not be defined there. From the second definition, if $p_a(t) = 0$, then $p_b(t) = 1$, and so:

$$p_{a\to b}(t', t) = \frac{P(S(t) = b \cap S(t') = a)}{P(S(t') = a)} = \frac{P(S(t') = a)}{P(S(t') = a)} = 1$$

Since if $P(B) = 1$, then $P(B \cap A) = P(A)$. Thus, $p_{a\to b}(t', t) = 1$ for $t' < t$ where we have assumed that $p_a(t') > 0$. (If $p_b(t) = 1$, i.e., the probabilities of the process guarantee $S = b$ at time $t$, then they must guarantee a transition to $b$ from a different state at earlier time $t'$.)

Another way to see this is to consider the law of total probability again. Taking $a = 0$ for concreteness, in the case where $p_0(t) = 0$, then $p_1(t) = 1$, and for $t' < t$:

$$1 = p_1(t) = p_{0\to1}(t', t)p_0(t') + p_{1\to1}(t', t)p_1(t')$$

Given that $p_1(t') = 1 - p_0(t')$, the only values between 0 and 1 that satisfy this equation are $p_{0\to1}(t', t) = p_{1\to1}(t', t) = 1$, for any non-zero $p_0(t')$, or $p_{1\to1}(t', t) = 1$ and $p_{0\to1}(t', t)$ unconstrained for $p_0(t') = 0$.

So $p_a(t) = 0$ implies $p_{a\to b}(t', t) = 1$. But then the limit diverges:

$$W_{a\to b}(t) = -\lim_{\delta t\to 0^-} \frac{p_{a\to b}(t + \delta t, t)}{\delta t} = -\lim_{\delta t\to 0^-} \frac{1}{\delta t} = +\infty$$

In [2] and [4], Gillespie argues that the divergence of $W_{a\to b}(t)$ violates the axioms of probability. The foregoing analysis demonstrates that the divergence of $W_{a\to b}(t)$ is in fact a consequence of the behaviour of the stochastic process, consistent with the axioms of probability, at times when $p_a(t) = 0$. Gillespie seemingly fails to properly consider this special case in [2] and [4] (though I do not have access to his textbook for more details on his argument).

It seems worth mentioning that the divergence of the probability rate at states with zero probability is analogous to the divergence of the Bohmian guiding velocity at configurations where the wavefunction is zero. In both cases, the problematic value (infinite jump rate, or infinite velocity) is never realized because the physical state never occupies the probability-zero configuration.

Consider the relationship between the single-time probabilities and the conditional probabilities:

$$p_a(t) = p_{0\to a}(0, t)p_0(0) + p_{1\to a}(0, t)p_1(0)$$

Taking time derivatives:

$$\frac{dp_a(t)}{dt} = \frac{\partial p_{0\to a}(0, t)}{\partial t}p_0(0) + \frac{\partial p_{1\to a}(0, t)}{\partial t}p_1(0)$$

Using the probability rate equations for the conditional probabilities, we find we can combine terms and apply the law of total probability again to arrive at:

$$\frac{dp_0(t)}{dt} = -W_{0\to1}(t)p_0(t) + W_{1\to0}(t)p_1(t)$$

$$\frac{dp_1(t)}{dt} = W_{0\to1}(t)p_0(t) - W_{1\to0}(t)p_1(t)$$

Thus, the single-time probabilities obey the same equations as the conditional probabilities, and as the quantum probabilities. This shows that if we want the relationship $p_a(t) = y_a(t)$, the jump rates for the process should be the same as the ones arrived at by the heuristic interpretation in the previous section. Of course, we must be cautious at moments when either of $p_0(t)$ or $p_1(t)$ becomes zero, but careful handling of those points is possible (e.g., as sketched at the end of the next section). On intervals between these points, the problem is well-behaved.

**Solving the Probability Rate Equation**

For the case of the two-state system, we can solve these equations and show that the probabilities are well-defined. Consider the probability of arriving at state 0 from a given earlier state:

$$\frac{\partial p_{a\to 0}(t_0,t)}{\partial t} = -W_{0\to 1}(t)p_{a\to 0}(t_0,t) + W_{1\to 0}(t)p_{a\to 1}(t_0,t)$$

We can substitute $p_{a\to 1}(t_0,t) = 1 - p_{a\to 0}(t_0,t)$ to write this as an uncoupled ODE:

$$\frac{\partial p_{a\to 0}(t_0,t)}{\partial t} + W_{tot}(t)p_{a\to 0}(t_0,t) = W_{1\to 0}(t)$$

Where:

$$W_{tot}(t) = W_{0\to 1}(t) + W_{1\to 0}(t)$$

Define the integrating factor for $0 \le t \le t'$:

$$\Phi(t,t') = \exp\left(-\int_t^{t'} W_{tot}(s)\, ds\right), \qquad \frac{\partial \Phi(t,t')}{\partial t} = W_{tot}(t)\Phi(t,t')$$

Note some properties of the integrating factor (assuming that $W_{1\to 0}(t) \ge 0$, $W_{0\to 1}(t) \ge 0$, and the integral of $W_{tot}$ exists on the relevant interval):

$$\Phi(t,t) = 1\,, \qquad 0 \le \Phi(t,t') \le 1\,, \qquad \Phi(t,t'') = \Phi(t,t')\Phi(t',t'')$$

Multiply the equation by the integrating factor and integrate:

$$\Phi(t,t_1)\frac{\partial p_{a\to 0}(t_0,t)}{\partial t} + W_{tot}(t)\Phi(t,t_1)p_{a\to 0}(t_0,t) = W_{1\to 0}(t)\Phi(t,t_1)$$

$$\frac{\partial}{\partial t}\big(\Phi(t,t_1)p_{a\to 0}(t_0,t)\big) = W_{1\to 0}(t)\Phi(t,t_1)$$

$$\Phi(t_1,t_1)p_{a\to 0}(t_0,t_1) - \Phi(t_0,t_1)p_{a\to 0}(t_0,t_0) = \int_{t_0}^{t_1} W_{1\to 0}(s)\Phi(s,t_1)\, ds$$

Performing the same analysis for $p_{a\to 1}(t_0, t)$, we get the following results:

$$p_{0\to 0}(t_0, t_1) = \Phi(t_0, t_1) + \int_{t_0}^{t_1} W_{1\to 0}(s)\Phi(s, t_1)\, ds$$

$$p_{1\to 0}(t_0, t_1) = \int_{t_0}^{t_1} W_{1\to 0}(s)\Phi(s, t_1)\, ds$$

$$p_{0\to 1}(t_0, t_1) = \int_{t_0}^{t_1} W_{0\to 1}(s)\Phi(s, t_1)\, ds$$

$$p_{1\to 1}(t_0, t_1) = \Phi(t_0, t_1) + \int_{t_0}^{t_1} W_{0\to 1}(s)\Phi(s, t_1)\, ds$$

All these expressions are greater than or equal to zero assuming that the probability rates are greater than or equal to zero, and moreover, the probabilities are bounded by one. For we can show that:

$$p_{0\to 0}(t_0, t_1) + p_{0\to 1}(t_0, t_1) = p_{1\to 0}(t_0, t_1) + p_{1\to 1}(t_0, t_1) = \Phi(t_0, t_1) + \int_{t_0}^{t_1} W_{tot}(s)\Phi(s, t_1)\, ds$$

But then the integral on the right evaluates to:

$$\int_{t_0}^{t_1} W_{tot}(s)\Phi(s, t_1)\, ds = \int_{t_0}^{t_1} \frac{\partial \Phi(s, t_1)}{\partial s} ds = \Phi(t_1, t_1) - \Phi(t_0, t_1) = 1 - \Phi(t_0, t_1)$$

Leaving us with the sums of the complementary probabilities equal to one:

$$p_{0\to 0}(t_0, t_1) + p_{0\to 1}(t_0, t_1) = p_{1\to 0}(t_0, t_1) + p_{1\to 1}(t_0, t_1) = 1$$

As a way of interpreting the integrating factor, consider that if $W_{a\to b}(t)dt$ is the probability of a jump from $S = a$ to $S = b$ in the infinitesimal interval $(t, t + dt)$, then the probability of no jump occurring from $t$ to $t'$ is the product integral:

$$P(\text{no jump in } (t, t') \mid S(t) = a) = \prod_{t}^{t'} (1 - W_{a\to b}(s)ds) = \exp\left(-\int_{t}^{t'} W_{a\to b}(s)\, ds\right)$$

The integrating factor $\Phi(t, t')$ is then the product of two probabilities that no jump occurs in $(t, t')$, respectively conditioned on $S(t) = 0$ and $S(t) = 1$. (This may also make us wary of the value of $\Phi(t, t')$ when one of the single-time probabilities $p_0(t)$ or $p_1(t)$ is zero.)

Let us now consider the case where we have the Hamiltonian given above, assuming the initial quantum state $|\psi(0)\rangle = |0\rangle$. Then by the quantum equilibrium hypothesis:

$$p_1(0) = y_1(0) = 0$$

So the conditional probabilities $p_{1\to0}(0,t)$ and $p_{1\to1}(0,t)$ are irrelevant. If we work with the other probabilities in the form:

$$p_{0\to1}(0,t) = \int_0^t W_{0\to1}(s)\Phi(s,t)\,ds$$

$$p_{0\to0}(0,t) = 1 - p_{0\to1}(0,t)$$

Using the jump rates calculated for this Hamiltonian:

$$W_{0\to1}(t) = \frac{\omega[\sin(\omega t)]^+ + 2J'(t)}{1+\cos(\omega t)}\ , \qquad W_{1\to0}(t) = \frac{\omega[-\sin(\omega t)]^+ + 2J'(t)}{1-\cos(\omega t)}$$

Then these expressions are defined and give sensible results, even for non-minimal choices of the probability flux $J' > 0$ (for which $W_{1\to0}$ diverges at $t = 0$), up to $t = \pi/\omega$ where $p_1(t)$ approaches 1 and $W_{0\to1}(t)$ diverges. Within that interval, the general expressions for two times, $0 < t_0 < t_1 < \pi/\omega$, are well behaved, with the conditional probabilities for $S(t) = 1$ approaching 1 as $t$ approaches $\pi/\omega$. At that endpoint, we can restart and use the probability $p_{1\to0}(\pi/\omega\,, t)$, and so on.

(Perhaps other Hamiltonians can result in jump rates that do not allow the integrals in the above expressions for the conditional probabilities to be evaluated. But at least in this case, the expressions work on intervals $t, t' \in (0, \pi/\omega)$, $t, t' \in (\pi/\omega\,, 2\pi/\omega)$, etc., and the Markov condition together with the extremal probabilities at those endpoints allows calculating all the other conditional probabilities for the process.)

**Addressing Criticisms**

In [2] and [4], Gillespie has several criticisms of the validity of the process proposed here. The first was already discussed briefly: he claims that the jump rates $W_{a\to b}(t)$ must always be bounded, or the conditional probabilities will go out of bounds (Gillespie's Theorem 1 in reference [2]). However, the preceding section shows that it is possible to have valid probabilities even with jump rates that diverge instantaneously, a direct counterexample to Gillespie's claim. (Again, it seems that the error in Gillespie's argument for Theorem 1 is in not properly handling cases where one of the single-time probabilities is zero.)

Second, he (essentially) claims that the rates in the model do not obey the definition:

$$W_{a\to b}(t) \equiv \lim_{\delta t\to0^+} \frac{p_{a\to b}(t, t+\delta t)}{\delta t} = \left.\frac{\partial p_{a\to b}(t,t')}{\partial t'}\right|_{t'=t}$$

But this is not true: for example, differentiating the following expression, it straightforwardly follows that the jump rate $W_{0\to1}(t)$ obeys this definition:

$$p_{0\to1}(t,t') = \int_t^{t'} W_{0\to1}(s)\Phi(s,t')\,ds \Rightarrow \left.\frac{\partial p_{a\to b}(t,t')}{\partial t'}\right|_{t'=t} = W_{0\to1}(t)$$

Again, I believe part of the disagreement may be arising from Gillespie attempting to apply this definition at times where the conditional probability is not necessarily defined. But here, another part of the disagreement comes from his confusion between:

$$P(S(t') = b \mid S(t) = a, |\psi(t)\rangle = |a\rangle)$$

Which is not the relevant probability in the Bohmian model, and:

$$p_{a\to b}(t,t') = P(S(t') = b \mid S(t) = a, |\psi(0)\rangle = |0\rangle)$$

Which is the relevant probability (with the conditioning on the initial quantum state made explicit). His equations 22a and 22b in [2] exhibit exactly this confusion; though this confusion is understandable since he is addressing Hardy et. al.'s model with its implicit "clock variable", rather than a Bohmian model.

Third, this confusion leads to his claim that the infinitesimal probability $W_{a\to b}(t)dt$ is quadratic rather than linear in $dt$; this is not true in general since the probabilities do not have the form that he assumes. However, a residue of this claim remains at times where one of the single-time probabilities $p_a(t)$ is zero, such as $t = 0$ if we assume $|\psi(0)\rangle = |0\rangle$. Then we must have:

$$p_{0\to1}(0,t) = \sin^2(\omega t/2)\ , \qquad W_{0\to1}(0) \equiv \lim_{\delta t\to0^+} \frac{p_{0\to1}(0,\delta t)}{\delta t} = \lim_{\delta t\to0^+} \left(\frac{\omega}{2}\right)^2 \delta t = 0$$

Then the jump rate is instantaneously 0, and moreover, for small times $\delta t$ the conditional probability $p_{0\to1}(0,\delta t)$ is quadratic in $\delta t$.

Let us first address one potential problem with $W_{0\to1}(0) = 0$: this nominally disagrees with the proposed jump rates if there is an excess probability flux $J'(0) > 0$. However, let us carefully consider the full conditional probability:

$$p_{0\to1}(0,t) = \int_0^t W_{0\to1}(s)\Phi(s,t)\,ds$$

From its construction, we already know this integral satisfies the first-order, linear ODE:

$$\frac{\partial p_{0\to1}(0,t)}{\partial t} + W_{tot}(t)p_{0\to1}(0,t) = W_{0\to1}(t)\ , \qquad p_{0\to1}(0,0) = 0$$

For times $0 < t < \pi/\omega$, the jump rates evaluate to:

$$W_{0\to1}(t) = \frac{\omega\sin(\omega t) + 2J'(t)}{1+\cos(\omega t)}\,, \qquad W_{1\to0}(t) = \frac{2J'(t)}{1-\cos(\omega t)}$$

$$W_{tot}(t) = W_{0\to1}(t) + W_{1\to0}(t) = \frac{\omega\sin(\omega t)}{1+\cos(\omega t)} + 2J'(t)\left(\frac{1}{1+\cos(\omega t)} + \frac{1}{1-\cos(\omega t)}\right)$$

One can verify that $\sin^2(\omega t/2)$ satisfies this ODE, no matter the value $J'(t)$, so by uniqueness of the ODE solution, $p_{0\to1}(0,t) = \sin^2(\omega t/2)$.

To summarize the third point so far, there seems to be an instantaneous breakdown in the definition of $W_{0\to1}(t)$ at $t=0$, which related to the fact that $p_1(0)=0$ and so, strictly speaking, the derivation of the probability rate equations also breaks down there. But this doesn't affect the actual conditional probability $p_{0\to1}(0,t)$, which is well-defined.

Still, Gillespie's concern with this third point is that it is inconsistent for $p_{0\to1}(t,t+\delta t)$ to be quadratic in $\delta t$, and in the model proposed here, $p_{0\to1}(t,t+\delta t)$ is approximately proportional to $(\delta t)^2$ *at certain times*. However, Gillespie's proof of the inconsistency (Theorem 2 in [2]) assumes that $p_{0\to1}(t,t+\delta t)$ is proportional to $(\delta t)^2$ *across a small interval* in $t$, rather than only instantaneously as in the Bohmian model. (Specifically, his equation 24 in [2] is not valid for the proposed model: the conditional probability has the assumed form at $t=0$, but *not* at $t=\delta t/2$.)

As an example for further clarity, taking $J'=0$, we can calculate in the proposed model that when both $\epsilon$ and $\delta t$ are small (using trapezoid rule for the integrals, Taylor expanding functions, and ignoring higher than second order terms in the small parameters):

$$W_{0\to1}(\epsilon) = W_{tot}(\epsilon) = \frac{\omega\sin(\omega\epsilon)}{1+\cos(\omega\epsilon)} \approx \frac{1}{2}\omega^2\epsilon$$

$$\Phi(\epsilon,\epsilon+\delta t) = \exp\left(-\int_{\epsilon}^{\epsilon+\delta t} W_{tot}(s)\,ds\right) \approx 1 - \frac{1}{2}\omega^2\epsilon\delta t - \frac{1}{4}\omega^2(\delta t)^2$$

So, the following conditional probability is approximately:

$$p_{0\to1}(\epsilon,\epsilon+\delta t) = \int_{\epsilon}^{\epsilon+\delta t} W_{0\to1}(s)\Phi(s,\epsilon+\delta t)\,ds \approx \frac{1}{2}\omega^2\epsilon\delta t + \frac{1}{4}\omega^2(\delta t)^2$$

Repeating Gillespie's argument for Theorem 2 with this correct expansion, no contradiction is found (the additional term restores the lost ½ on one side of the equation). Thus, there is no inconsistency in the proposed jump rates for the Bohmian model leading to times where $p_{0\to1}(t,t+\delta t)$ is instantaneously proportional to $(\delta t)^2$. These are just times where the jump rate $W_{0\to1}(t)$ is instantaneously equal to zero and increasing from there.

## Conclusion

The above analysis demonstrates that there is no apparent inconsistency in a Bohmian version of a quantum system with 2 states; the probabilities for the stochastic evolution of the physical state are all well-defined in the considered model. Of course, the focus on a two-state system doesn't immediately prove that this kind of Bohmian system is valid in general, but there does not seem to be any reason to expect qualitatively different behaviour for systems with more than 2 states.

## Acknowledgement

Many thanks to Jacob Barandes for insightful discussion, and for critiquing the first version of this document.

## References

[1] Hardy, L., Home, D., et al. Realism and the quantum-mechanical two-state oscillator. *Phys. Rev. A.* **45** 7 (1992). https://doi.org/10.1103/PhysRevA.45.4267

[2] Gillespie, D. Why quantum mechanics cannot be formulated as a Markov process. *Phys. Rev. A.* **49** 3 (1994). https://doi.org/10.1103/PhysRevA.49.1607

[3] Hardy, L., Home, D., et al. Comment on 'Why quantum mechanics cannot be formulated as a Markov process'. *Phys. Rev. A.* **56** 4 (1997). https://doi.org/10.1103/PhysRevA.56.3301

[4] Reply to "Comment on 'Why quantum mechanics cannot be formulated as a Markov process'". *Phys. Rev. A.* **56** 4 (1997). https://doi.org/10.1103/PhysRevA.56.3304

[5] Robinson Erhardt Podcast. "David Albert & Jacob Barandes: Debating the Foundation of Quantum Mechanics". YouTube

[6] Bell, J.S. "Beables for quantum field theory" in *Speakable and Unspeakable in Quantum Mechanics*, pp. 173-180. Cambridge University Press (2011). https://doi.org/10.1017/CBO9780511815676

[7] Durr, D., Goldstein, S., Tumulka, R., Zanghi, N. Bell-Type Quantum Field Theories. *J. Phys. A.* **38** R1 (2005). https://doi.org/10.1088/0305-4470/38/4/R01

[8] Dickau, M. Apparent energy-speed relationship poses no challenge to Bohmian mechanics (2025). https://arxiv.org/abs/2507.18826